\documentstyle{article}
\begin{document}

\renewcommand{\arraystretch}{1.0}
\renewcommand{\textfraction}{0.05}
\renewcommand{\topfraction}{0.9}
\renewcommand{\bottomfraction}{0.9}
\def\la{\;\raise0.3ex\hbox{$<$\kern-0.75em\raise-1.1ex\hbox{$\sim$}}\;}
\def\ga{\;\raise0.3ex\hbox{$>$\kern-0.75em\raise-1.1ex\hbox{$\sim$}}\;}
\def\lr{\;\raise0.3ex\hbox{$\rightarrow$\kern-1.0em\raise-1.1ex\hbox{$\leftarrow$}}\;}
\newcommand{\kB}{\mbox{$k_{\rm B}$}}           
\newcommand{\tn}{\mbox{$T_{{\rm c}n}$}}        
\newcommand{\tp}{\mbox{$T_{{\rm c}p}$}}        
\newcommand{\te}{\mbox{$T_{eff}$}}             
\newcommand{\dash}{\mbox{--}}                  
\newcommand{\ex}{\mbox{\rm e}}                 
\newcommand{\r}{\rule{0cm}{0.4cm}}
\newcommand{\hh}{\rule{0.5cm}{0cm}}
\newcommand{\hb}{\rule{0.4cm}{0cm}}
\newcommand{\rl}{\rule{0em}{1.8ex}}
\newcommand{\hhh}{\rule{1.4em}{0ex}}
\newcommand{\hhb}{\rule{1.2em}{0ex}}
\newcommand{\hhl}{\rule{2.4em}{0ex}}
\newcommand{\s}{$\;\;$}

\title{\bf High-energy accelerators above pulsar polar caps}

\author{R.X.~Xu, G.J.~Qiao, B.~Zhang\\
        CAS-PKU joint Beijing Astrophysical Center
        and Department of Astronomy,\\
\vspace{3mm}
        Peking University, Beijing 100781, China\\
\vspace{3mm}
{\it Presented in IAU Colloquium 177: Pulsar Astronomy and Beyond}\\
        {\it August 30 -- Setpember 3, 1999, Bonn, Germany}
}
\date{}

\maketitle

\begin{abstract}

Similar to the terrestrial collision accelerators of e$^{\pm}$,
another kind of accelerator is above a
positively or negatively charged pulsar polar cap. In the case of
pulsars with magnetic axis parallel (anti-parallel) to rotational
axis, relativistic e$^+$ (e$^-$) with Lorentz factor
$\gamma\sim 10^6$ hit the electrons in the polar caps. These scenarios
are investigated both for pulsars being BSSs (bare strange stars) and
for pulsars being NSs (neutron stars). Such a study may be valuable to
differentiate NSs and BSSs observationally.

\end{abstract}

\vspace{5mm}
\noindent
%
High-energy electron-positron collision in laboratory, where the
energy in the center of mass ($E_{\rm cm}$) is about GeV, has been
successfully used to study the structure of elementary particles and
their interactions. For anti-pulsars, when a backflow
e$^+$ beats an electron in polar cap, $E_{\rm cm}$ is {\it also}
about GeV. $E_{\rm cm}=0.511~{\rm MeV}~\sqrt{2\gamma}\sim1$ GeV for
$\gamma\sim 10^6$. Therefore, the physics in the polar cap
accelerators does not go beyond that in laboratory.

For rotation-powered pulsars, the rotation energy is converted into
the kinetic and radiative energy via an induced electric field around
a magnetized rotator, hence, copious
e$^{\pm}$ in the open field lines will be accelerated in different
directions. The outflow produces a power law radiation, while the
backflow heats the polar cap. Therefore, nearly half the loss of
rotation energy,
$
\dot E = -9.6\times 10^{30} P_1^{-4}R_6^6B_{12}^2
$ ergs/s
, is in outflow and backflow. In fact, relativistic backflow particles
radiate away part of their energy before reaching the cap. Considering
the polar gap and the out gap accelerations, one can find the residual
energy of a charged particles striking the cap$^1$ being
$\sim 5.9$ ergs (thus $\gamma\sim 10^6$).
With a Goldreich-Julian current bombardment,
the polar cap will receive an energy rate$^1$
$
E_{\rm rate} \sim 8.1\times 10^{30} P_1^{-5/3}B_{12}R_6^3
$ ergs/s, which is in order of $\dot E$. The
cap has a radius of
$
r_{\rm p}=1.45\times 10^4R_6^{3/2}P_1^{-1/2}
$ cm, thus the energy flux received is
$
F_{\rm p} = 1.8\times 10^{22}B_{12}P_1^{-2/3}
~{\rm ergs} \cdot {\rm s}^{-1} \cdot {\rm cm}^{-2}
$.

It was believed that NSs and SSs (strange stars) can be
differentiated observationally
by studying the global cooling behaviors. However, it is found that
SSs cool significantly more rapidly than NSs within the first
$\sim$30 yr after birth$^2$, which makes NSs and SSs almost
un-distinguishable based only on the global thermal radiation.
Now that pulsars may be BSSs (see, e.g., Xu, Qiao, \& Zhang
in this proceeding), and the materials in the polar caps of
NSs and BSSs are very different, we suggest to study the {\it
local thermal behavior in the polar caps} in order to distinguish NSs
and SSs. Such suggestion avoids involving us in the detailed
microphysics processes in the interiors of NSs and SSs.


\vspace{5mm}
\noindent
{\bf BSSs: have cooler polar caps?}~~
As a polar cap is hotter than the other part in a pulsar's surface,
heat flows from the cap to the equator. If pulsars are NSs, such heat
flow is negligible, and an electromagnetic shower produced by
incident particles can almost be converted into thermal radiation
re-radiated. If there is no thermal conductivity, the temperature in
the cap is
$
T_0=({F_{\rm p}\over \sigma})^{1/4}
\sim 4.22\times 10^6
$K. The coefficient of
thermal conductivity mainly due to the transport of heat by electrons
in NS surface is given by$^3$
$
\kappa^{\rm NS} = 3.8\times 10^{14} \rho_5^{4/3}
~{\rm ergs}\cdot {\rm s}^{-1}\cdot {\rm cm}^{-1}\cdot {\rm K}^{-1}
$ ($\rho_5$ is the density in unit of $10^5$ g cm$^{-3}$),
which are nearly independent of the details of the lattice. The
temperature gradient in the crust of a NS is $\nabla T^{\rm NS}\sim
{T_0\over r_{\rm p}}$, then the heat flow
$
H^{\rm NS}\sim \kappa^{\rm NS} \nabla T^{\rm NS} r_{\rm p}^n
\sim \kappa^{\rm NS} T_{\rm 0} \sqrt{r_{\rm p}}
\sim 1.9\times 10^{23}
$ ergs/s $\ll E_{\rm rate}$ ($n=1\sim 2$ when the thermal conduction
geometry near polar cap is considered, we let $n=1.5$ here).

However, if pulsars are BSSs, the electron number density in the
surface of a BSS,
$
n^{\rm BSS} = 1.5\times 10^{34} {\rm cm}^{-3}
$
, is much lager than that of an NS,
$
n^{\rm NS} = 2.8\times 10^{28} \rho_5 {\rm cm}^{-3}
$.
Hence, the coefficient of thermal conductivity in BSS surface,
$
\kappa^{\rm SS}\sim ({n^{\rm SS}\over n^{\rm NS}})^{4/3}\kappa^{\rm
NS}
\sim 4.35\times 10^7\kappa^{\rm NS}
$, although there is no detailed calculation of $\kappa^{\rm SS}$ in
literature. Therefore the correspondent heat flow in BSS,
$
H^{\rm SS}\sim 8.3\times 10^{30}
$ ergs/s
, is comparable with $E_{\rm rate}$, being un-negligible. Such
heat flow will result in {\it a much lower polar cap temperature} in bare
strange stars.


%
\vspace{5mm}
\noindent
{\bf Anti-pulsars with ${\bf \Omega\cdot B} > 0$}~~
Electrons are pulled out, while positrons bombard the cap. Such process
is very similar to that in the laboratory collision of e$^{\pm}$:
${\rm e}^+{\rm e}^-\rightarrow {\rm f}\overline{\rm f}({\rm e}^+{\rm
e}^-), \gamma\gamma$ (${\rm f}\overline{\rm f}$ is a pair of fermion).
A gauss-like peak $\sim$0.511 MeV spectrum could be observed. However,
if pulsars are NSs, additional bremsstrahlung may produces a lower
continue spectrum superposed on the peak spectrum.
Therefore, NSs and
BSSs may be distinguished by spectrum observations.

\vspace{5mm}
\noindent
{\bf Pulsars with ${\bf \Omega\cdot B} < 0$}~~
Positive charged particles are pulled out, while an electron shower
pours on the cap. If pulsars are NSs, the pulled ions being Fe and/or
He nucleus
(the composition of the cap surface is uncertain)
might result in a formation of line spectrum in 0.04 keV $\sim$ 10
keV according to the energy levels of atom $E_n = -13.6\;{\rm eV}\;\;
Z^2 {1\over n^2}$. No such line can be observed if pulsars are BSSs.
Observationally, there is no convinced signature of ion line in
rotation-powered pulsars. Hence, such pulsars may tend to be BSSs.
Perhaps, the future missions of Astro-E
(with a fine resolution spectroscope in 0.4 - 10 keV) would
make it clear, that heavy ion (such as iron) lines
can or cannot
be seen from rotation-powered pulsars.

\vspace{5mm}
\noindent
{\bf Conclusion}~~
Two conclusions:
1. The polar cap of bare strange star may be cooler than that of
neutron star;
2. Spectrum observation might tell that a pulsar is a bare strange
star or a neutron star.

\vspace{7mm}
We thank Dr. J.L. Han, Mr. B.H. Hong, and other members in our
pulsar group.
This work is supported by NSFC (No. 19803001,
N0.19910211260-570-A03), by the Climbing project of China, by
Doctoral Program Foundation of Institution of Higher Education
in China and by the Youth Foundation of PKU.

\begin{quote}

\verb"[1] Wang, F., et al. 1998, ApJ, 498, 373"\\
\verb"[2] Schaab, C., et al. 1997, ApJ, 480, L111"\\
\verb"[3] Jones, P. B. 1978, MNRAS, 184, 807"

\end{quote}

\end{document}